\documentclass[conference]{IEEEtran}
\IEEEoverridecommandlockouts
\usepackage{cite}
\usepackage{amsmath,amssymb,amsfonts,amsthm}
\usepackage{graphicx}
\usepackage{textcomp}
\usepackage{xcolor}

\usepackage{multirow}
\usepackage{makecell}
\usepackage{algpseudocode}
\usepackage{booktabs}
\usepackage{colortbl}
\usepackage{algorithm}
\usepackage{array}
\usepackage{stfloats}
\usepackage{url}
\usepackage{verbatim}
\usepackage{setspace}
\newtheorem{definition}{{Definition}}

\usepackage{color, xspace}

\def\BibTeX{{\rm B\kern-.05em{\sc i\kern-.025em b}\kern-.08em
    T\kern-.1667em\lower.7ex\hbox{E}\kern-.125emX}}
\begin{document}
\title{Multi-triplet Feature Augmentation for Ponzi Scheme Detection in Ethereum\\

}

\author{
    \IEEEauthorblockN{
        Chengxiang Jin\IEEEauthorrefmark{1}\IEEEauthorrefmark{2},
        Jiajun Zhou\IEEEauthorrefmark{1}\IEEEauthorrefmark{2},
        Shengbo Gong\IEEEauthorrefmark{1}\IEEEauthorrefmark{2}, 
        Chenxuan Xie\IEEEauthorrefmark{1}\IEEEauthorrefmark{2}, and
        Qi Xuan\IEEEauthorrefmark{1}\IEEEauthorrefmark{2},~\IEEEmembership{Senior Member,~IEEE}}
    \IEEEauthorblockA{
        \IEEEauthorrefmark{1}Institute of Cyberspace Security, Zhejiang University of Technology, Hangzhou, China}
    \IEEEauthorblockA{
        \IEEEauthorrefmark{2}Binjiang Cyberspace Security Institute of ZJUT, Hangzhou, China}
    \IEEEauthorblockA{
        \{jincxiang, jjzhou, jshmhsb, 221122030330, xuanqi\}@zjut.edu.cn
    }
    \thanks{This work was supported in part by 
            the Key R\&D Program of Zhejiang under Grant 2022C01018, 
            the National Natural Science Foundation of China under Grant 62103374,
            the National Natural Science Foundation of China under Grants U21B2001, 
            the National Key R\&D Program of China under Grant 2020YFB1006104.
            \textit{Corresponding author: Jiajun Zhou (email: jjzhou@zjut.edu.cn)}.}
}

\maketitle

\begin{abstract}
Blockchain technology revolutionizes the Internet, but also poses increasing risks, particularly in cryptocurrency finance. On the Ethereum platform, Ponzi schemes, phishing scams, and a variety of other frauds emerge. Existing Ponzi scheme detection approaches based on heterogeneous transaction graph modeling leverages semantic information between node (account) pairs to establish connections, overlooking the semantic attributes inherent to the edges (interactions). To overcome this, we construct heterogeneous Ethereum interaction graphs with multiple triplet interaction patterns to better depict the real Ethereum environment. Based on this, we design a new framework named multi-triplet augmented heterogeneous graph neural network (MAHGNN) for Ponzi scheme detection. We introduce the Conditional Variational Auto Encoder (CVAE) to capture the semantic information of different triplet interaction patterns, which facilitates the characterization on account features. Extensive experiments demonstrate that MAHGNN is capable of addressing the problem of multi-edge interactions in heterogeneous Ethereum interaction graphs and achieving state-of-the-art performance in Ponzi scheme detection.
\end{abstract}


\section{Introducing}

Ponzi schemes, identified as a form of fraudulent investment scheme~\cite{artzrouni2009mathematics}, have witnessed an escalating prevalence within Ethereum~\cite{wood2014ethereum}. Typically enticing investors with promises of substantial returns, these schemes operate by redistributing investments from new participants rather than generating profits through legitimate business activities. Such deceptive practices invariably collapse when there is an insufficient influx of new investors to sustain payouts for previous participants, resulting in substantial financial losses for those involved.
On November 4th, 2022, the United States Securities and Exchange Commission (SEC) filed charges against Trade Coin Club for operating a fraudulent cryptocurrency Ponzi scheme and raising \$295 million\footnote[1]{\url{https://www.sec.gov/news/press-release/2022-201}}. 
The SEC further alleges that investor withdrawals are solely funded by new investor deposits, indicating a consistent pattern of a Ponzi scheme. Therefore, the detection of Ponzi schemes in Ethereum is an urgent and crucial matter requiring immediate attention.

Existing methods for detecting Ponzi schemes in Ethereum typically focus on constructing code-level features~\cite{bartoletti2020dissecting} and transaction-level features~\cite{jung2019data}, which are then combined with machine learning methods or graph algorithms for detection.
Code-level features often rely on statistical opcode characteristics, which may not be universally accessible for all contract codes and suffer from a cumbersome extraction process. Additionally, existing graph approaches tend to model Ethereum data as homogeneous graphs, disregarding the distinct roles of accounts and transactions, thereby leading to information loss and an inadequate depiction of actual account interaction patterns. 
The aforementioned issues constrain the detection performance of Ponzi schemes in Ethereum.
To address these limitations, various detection approaches based on heterogeneous graph modeling~\cite{shi2016survey} have emerged. 
These approaches consider the diverse types of both accounts and transactions, thereby improving the performance of account representation learning and Ponzi scheme detection. 
However, in most cases, these heterogeneous graph modeling approaches can only depict a single type of interaction between different node types. Conversely, within Ethereum, multiple types of interactions occur even between two specific account types, such as Ether transfers and contract calls between externally owned accounts and contract accounts.

In this regard, to better characterize complex Ethereum interaction scenarios, we propose a Heterogeneous Graph Neural Network method based on Multi-triplet Feature Augmentation (MAHGNN), which can effectively capture multiple triplet interaction patterns between target accounts and their surrounding accounts, enabling better characterization of the complex behavioral patterns exhibited by target accounts and facilitating powerful detection of Ponzi schemes.
The main contributions of this work are summarized as follows:
\begin{itemize}
    \item We construct a heterogeneous Ethereum interaction graph that contains multiple triplet interaction patterns.
    \item We propose a Multi-triplet Augmented Heterogeneous Graph Neural Network (MAHGNN), which can enrich the characteristics of target accounts by simulating the complex multi-triplet interactions around them.

    


    \item Extensive experiments show that MAHGNN consistently outperforms existing Ponzi scheme detection methods.
\end{itemize}

The rest of this paper is summarized as follows. 
Sec.~\ref{sec:related work} provides a review of previous work on Ethereum scheme detection. 
Sec.~\ref{sec:pre} describes the details of heterogeneous Ethereum interaction graph. 
Sec.~\ref{sec:method} introduces the details of proposed MAHGNN method. 
Sec.~\ref{sec:exp} presents the experimental setup and result analysis. 
Finally, Sec.~\ref{sec:Conclusion} concludes this paper and provides an outlook.

\section{Related Work} \label{sec:related work}
Fraud detection in blockchain continues to attract significant attention, with detection methods advancing from manual feature engineering to more powerful transaction graph analysis.

Chen et al.~\cite{chen2018detecting} constructed manual features for accounts and then fed them into downstream machine learning models to identify Ponzi schemes.
However, this approach heavily relies on expert knowledge and can only target schemes that conform to the pre-established transaction characteristics. 
To accurately capture the interaction patterns among accounts in transactions, graph embedding techniques based on random walk have been extensively employed. For instance, Wu et al.~\cite{wu2020phishers} proposed a novel network embedding algorithm called trans2vec, which leverages transaction amounts and timestamps within an Ethereum transaction graph.
Similarly, Tan et al.~\cite{tan2021graph} employed the transaction amounts on edges to compute walking probabilities using Node2vec~\cite{grover2016node2vec}, and subsequently utilized the resulting embeddings in downstream detection.
However, the aforementioned methods primarily focus on extracting information from the graph structure and do not effectively utilize the valuable node features. 
With the emergence of graph neural networks (GNNs), several detection algorithms based on graphs have been developed to capture both node features and structural information simultaneously. 
Yu et al.~\cite{yu2021ponzi} initially employed a graph convolutional network (GCN)~\cite{kipf2016semi} for identifying Ethereum Ponzi schemes, yielding promising results. 
However, the proposed approach relies on manual features as input, which may impose limitations on its expressiveness. Therefore, Tan et al.~\cite{tan2023ethereum} employ a graph embedding method to generate embeddings as input, followed by the utilization of GCN for fraud detection.


All the aforementioned methods are based on homogeneous graph modeling, overlooking the importance of account type and interaction type. In an early attempt to introduce heterogeneity in Ponzi detection, Jin et al.~\cite{jin2022heterogeneous} enhanced the existing homogeneous Ponzi detection technique by constructing heterogeneous graphs and employing metapaths. Concurrently, anomaly detection algorithms that leverage heterogeneous graphs emerge.
Based on heterogeneous embedding techniques, Wang et al.~\cite{wang2022heterogeneous} employed a biased random walk to acquire the embedding representation using transaction amounts and timestamps, and utilized a normalized heterogeneous softmax function based on node type. However, similar to homogeneous embedding methods, the heterogeneous embedding approach also exhibits certain limitations. Consequently, there is a growing trend towards adopting heterogeneous neural networks that offer promising solutions to address these drawbacks.
Liu et al.~\cite{liu2022blockchain} employed the transformer network for acquiring the paths linking multi-hop connected nodes and generated a metapath correlation matrix, which is subsequently fed into a convolutional neural network to procure the node embedding.


\section{Ethereum Interaction Graph Modeling} \label{sec:pre}
In this section, we mainly introduce the Ethereum data and the construction of heterogeneous Ethereum interaction graph.


\subsection{Ethereum Data} \label{subsec:data}
In Ethereum, an account represents an entity that holds Ether and can be classified into two types: Externally Owned Accounts (EOAs) and Contract Accounts (CAs). 
EOAs are managed by their respective private key holders, who have the capability to initiate transactions on the Ethereum network. 
CAs are governed by their underlying smart contract code and can only be triggered to execute functions defined within the contract.
Interactions between Ethereum accounts can be classified into two categories: transactions (\textit{trans}) and contract calls (\textit{call}).
Transactions primarily involve the transfer of Ether. Contract calls obtain various services by triggering functions within the smart contract. Further analyzing these interactions can provide deeper insights into the functioning of the Ethereum network and uncover risks.
    

\subsection{Heterogeneous Ethereum Interaction Graph} \label{sec:data}
Heterogeneous graphs encompass diverse types of edges and nodes, serving as an effective means to represent complex interactive systems in reality. 
However, in existing heterogeneous graph datasets such as ACM~\cite{zhao2020network}, DBLP~\cite{fu2020magnn} and IMDB~\cite{yu2022multiplex}, the types of edges between two specific types of nodes are also determined.
Taking ACM as an example, there are three types of nodes: author (A), paper (P) and subject(S), but only two types of edges: author of the paper (P-A) and subject of the paper (P-S).
However, in Ethereum, there will be multiple types of interactions even between two specific account types.
Here we construct heterogeneous Ethernet interaction graphs to further illustrate this difference.

\begin{definition}[\textbf{Heterogeneous Ethereum Interaction Graph, HEIG}]\label{def:HEIG}
    We treat Ethereum accounts as nodes and interactions between accounts as edges, constructing a Heterogeneous Ethereum Interaction Graph (HEIG), symbolized as $G=(V_\textit{eoa}, V_\textit{ca}, E_\textit{trans}, E_\textit{call}, Y)$, where $V_\textit{eoa}$ and $V_\textit{ca}$ represent the sets of EOAs and CAs respectively, $E_\textit{trans}$ and $E_\textit{call}$ represent the sets of transactions and contract calls respectively, and $Y = \{(v_i,y_i) \mid v_i \in V_\textit{ca}\}$ represents the label set of partial CA nodes with known identity information.
\end{definition}




\begin{figure}[t]
    \centering
    \includegraphics[width=\linewidth]{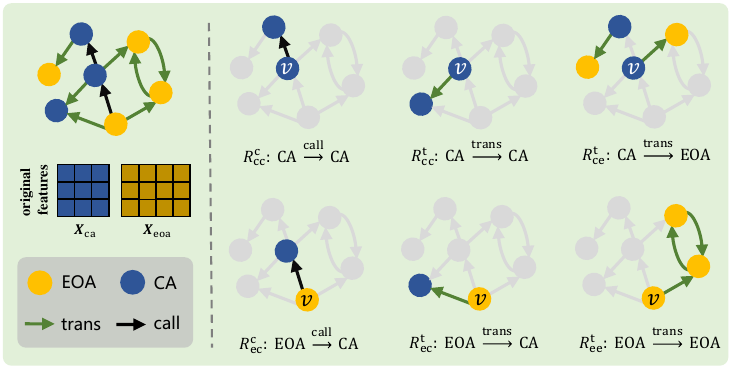}
    \caption{The framework of heterogeneous Ethereum interaction graph including multiplex triplet subgraphs.}
    \label{fig:graph}
\end{figure}

Based on this definition, we further discuss the different interaction patterns in Ethereum. According to the rules of Ethereum, the target of a \textit{call} edge must be a CA, while the source and target of a transaction edge are not restricted. Fig.~\ref{fig:graph} illustrates the multiple triplets formed by different account pairs and different types of interactions in HEIG.
Through a comprehensive analysis of account interactions in Ethereum, we can derive six triplet interaction patterns as follows:
\begin{equation} \label{eq: triples}
    \begin{aligned}
        &R_\textit{cc}^\textit{c}: \textit{CA} \stackrel{\text{call}}{\longrightarrow }\textit{CA},   \quad\ \   R_\textit{cc}^\textit{t}: \textit{CA}  \stackrel{\text{trans}}{\longrightarrow}\textit{CA}\\
        &R_\textit{ce}^\textit{t}: \textit{CA} \stackrel{\text{trans}}{\longrightarrow}\textit{EOA},  \quad      R_\textit{ec}^\textit{c}: \textit{EOA} \stackrel{\text{call}}{\longrightarrow} \textit{CA}\\
        &R_\textit{ec}^\textit{t}: \textit{EOA}\stackrel{\text{trans}}{\longrightarrow}\textit{CA},   \quad      R_\textit{ee}^\textit{t}: \textit{EOA} \stackrel{\text{trans}}{\longrightarrow}\textit{EOA}
    \end{aligned}
\end{equation}

\subsection{Account Feature Initialization}
When we apply graph-related algorithms, especially graph neural networks, to analyze HEIG, the initial node (account) features are indispensable. 
However, the account interaction graph defined in Definition~\ref{def:HEIG} is devoid of account features. 
Therefore, in this paper, we construct manual features for accounts to serve as their initial features. 
Specifically, for each account $v_i$ (CA or EOA), we construct manual features based on different interactions (\textit{trans} or \textit{call}) as follows:
\begin{itemize}
    \item Investment and returns generated under specific interaction types (including total and average, a total of $2 \times 2\times 2 = 8$ types).
    \item Balance obtained under specific interaction types (a total of $2\times 1\times 1=2$ types).
    \item Number of initiations and receptions under specific interaction types (a total of $2\times 2\times 1=4$ types).
\end{itemize}
The 14 manual features defined above are widely used to characterize the transaction features of accounts in blockchain.
After statistical computation, we use them to construct a 14-dimensional initial feature vector $\boldsymbol{X}_i \in \mathbb{R}^{14}$ for each account $v_i$ in HEIG.

\begin{figure}[t]
    \centering
    \includegraphics[width=0.8\linewidth]{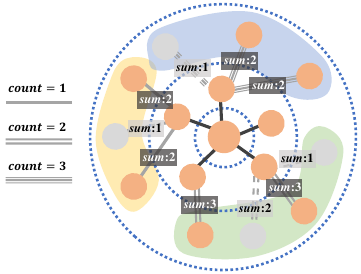}
    \caption{The framework for second-order neighborhood filtering based on edge features.}
    \label{fig:dataset}
\end{figure}

\subsection{Dataset Construction}

We retrieve a total of 191 labeled Ponzi accounts and 1,152 non-Ponzi accounts from various Blockchain data platforms, including \textit{Xblock}\footnote[1]{\url{http://xblock.pro/ethereum/}}, \textit{Etherscan}\footnote[2]{\url{https://cn.etherscan.com/accounts/label/ponzi}}.
Since the raw data in Ethereum is so massive, we scale down the data by filtering the relatively unimportant second-order neighbors of the target accounts.
Remarkably, edges have \textit{count} and \textit{sum} features. The former is the number of occurrences indicating the interaction frequency, the latter is the total amount indicating the size of the interaction.
We categorize the second-order neighbor data by the \textit{count} and subsequently retain \textit{top-k}\% of edges within each group that possess higher \textit{sum} features, as illustrated in Figure~\ref{fig:dataset}, thus ensuring attention to edges involving higher values and retaining the nearly original structural features.

We randomly sample 191 accounts from the labeled non-Ponzi accounts three times, and then filter their neighbors with \textit{k=0.01} and \textit{k=0.001} as the above method. As a result, we obtain six different datasets, as specified in Table~\ref{tab:data}. As an example, the filtered graph of Version-0 contains only 5\% CA of the corresponding raw data.
\begin{table}[t]
    \centering
    \caption{Statistics for different datasets with different \textit{top-k} sampling.}
        \resizebox{\linewidth}{!}{
            \renewcommand\arraystretch{1.3}
            \begin{tabular}{c|c|cccccccc} 
                \hline\hline
                Datasets                    & \textit{k}  & CA & EOA & $R_\textit{cc}^\textit{c}$ & $R_\textit{cc}^\textit{t}$ & $R_\textit{ce}^\textit{t}$ & $R_\textit{ec}^\textit{c}$ & $R_\textit{ec}^\textit{t}$ & $R_\textit{ee}^\textit{t}$  \\ 
                \hline
                \multirow{2}{*}{Version-0}  & 0.01  & 72,721    & 850,745    & 71,254   & 4,679  & 13,316  & 774,329   & 401,026  & 99,960   \\
                                            & 0.001 & 65,412    & 696,721    & 65,559   & 4,669  & 12,538  & 738,983   & 398,545  & 10,061   \\ 
                \hline
                \multirow{2}{*}{Version-1}  & 0.01  & 109,777   & 796,556    & 102,022  & 511    & 13,316  & 679,870   & 135,680  & 96,457   \\
                                            & 0.001 & 100,089   & 616,276    & 101,798  & 501    & 12,538  & 662,329   & 133,323  & 9,708   \\ 
                \hline
                \multirow{2}{*}{Version-2}  & 0.01  & 63,373    & 738,356    & 57,759   & 1,142  & 13,316  & 656,823   & 353,638  & 101,467  \\
                                            & 0.001 & 53,333    & 585,485    & 53,841   & 1,132  & 12,538  & 621,501   & 351,121  & 10,212   \\
                \hline\hline
                \end{tabular}}
    \label{tab:data}
\end{table}

\section{Methodology} \label{sec:method}
\begin{figure*}[t]
\centering
\includegraphics[width=\textwidth]{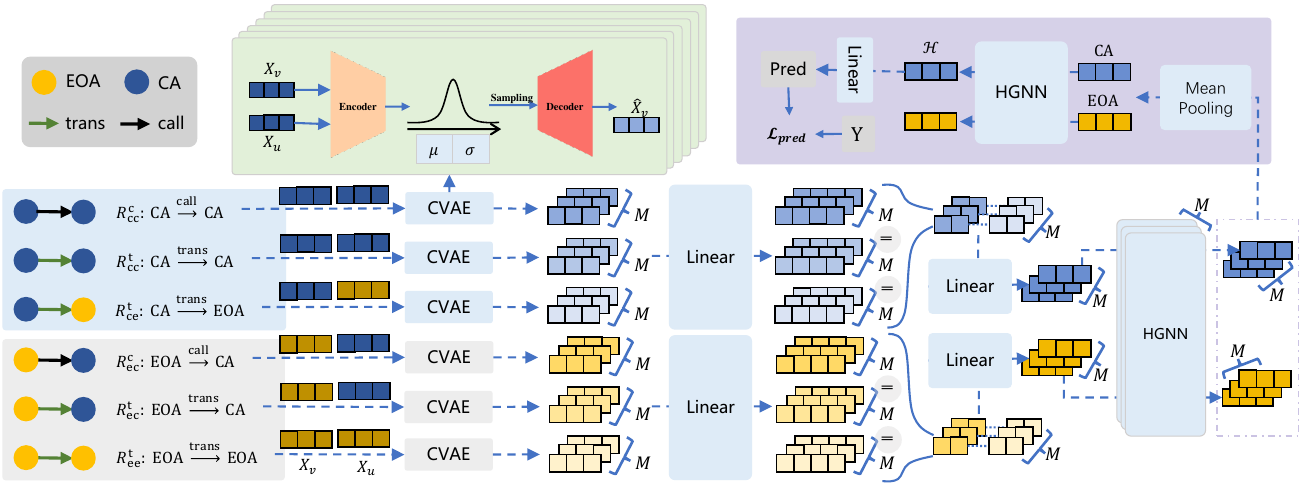}
\caption{The framework of Multi-triplet Augmented Heterogeneous Graph Neural Networks. The CVAE model, tailored to multi-triplet interactions, is utilized to generate the corresponding triplet-level augmentation features. 
Following the model training process, the prediction loss ($\mathcal{L}_{pred}$) is obtained.}
\label{fig:HLA}
\end{figure*}

In this section, we introduce the MAHGNN model, which utilizes multi-triplet interaction patterns to enhance the characterization of target accounts, 
The main framework is illustrated in Fig.~\ref{fig:HLA}.

\subsection{Triplet-level Information Generation Model}
%

For HEIG, the neighborhood of each account exhibits a diverse range of triplet interaction patterns, which are primarily influenced by the target account. To effectively capture these distribution features and provide insights into the behavioral patterns of the target node, we propose a triplet-level information generation model.

Specifically, our triplet-level information generation model can be regarded as a feature augmentation strategy,
which can generate more neighborhood features based on triplet starting from target account, thereby helping to improve the expressive power of Ponzi detection model.
Specifically, we first consider the initial features of target account ($\boldsymbol{X}_v$) as the conditional information and the initial features of all neighbor accounts ($\boldsymbol{X}_u$) as the input data.
Then we utilize a conditional variational auto-encoder (CVAE)~\cite{sohn2015learning,liu2022local} to takes the conditional information and input data to generate new neighborhood features for the target account. 
The CVAE consists of two parts: an encoder and a decoder.
The encoder learns how to map the input data $\boldsymbol{X}_u$ and conditional information $\boldsymbol{X}_v$ to the latent space, and outputs the distribution parameters of the latent variable $\boldsymbol{z}$:
\begin{equation}
  \boldsymbol{\mu}, \boldsymbol{\sigma} = \text{Encoder} (\boldsymbol{X}_v, \boldsymbol{X}_u)
\end{equation} 
where the mean $\boldsymbol{\mu}$ and standard deviation $\boldsymbol{\sigma}$ are used to describe the posterior probability distribution $q_\phi\left(\mathbf{z} \mid \boldsymbol{X}_u, \boldsymbol{X}_v\right)$.
The decoder takes the sampled latent variable $\mathbf{z}$ and conditional information $\boldsymbol{X}_v$ as input and generates features for the target node $v$:
\begin{equation}
  \begin{aligned}
    \boldsymbol{z} &= \boldsymbol{\mu} + \boldsymbol{\epsilon}  \odot  \boldsymbol{\sigma}\\
    \hat{\boldsymbol{X}}_v &= \text{Decoder}(\mathbf{z}, \boldsymbol{X}_v) 
  \end{aligned}
\end{equation}
where $\boldsymbol{\epsilon} \sim \mathcal{N}(\mathbf{0}, \mathbf{I})$ and $\odot$ is the element-wise multiplication.

During the training phase, the goal of CVAE is to learn the neighborhood information distribution of the target node using neighbor pairs $(\boldsymbol{X}_v, \boldsymbol{X}_u, u \in \mathcal{N}_v)$ as input, which is achieved by maximizing the evidence lower bound (ELBO) as follows:
\begin{equation}
  \begin{aligned}
  \mathcal{L}\left(\boldsymbol{X}_u, \boldsymbol{X}_v ; \theta, \phi\right)= & -\text{KL}\left(q_\phi(\boldsymbol{z} \mid \boldsymbol{X}_u, \boldsymbol{X}_v) \parallel  p_\theta(\boldsymbol{z} \mid \boldsymbol{X}_v)\right) \\
  & +\frac{1}{L} \sum_{l=1}^L \log p_\theta(\boldsymbol{X}_u \mid \boldsymbol{X}_v, \boldsymbol{z})
  \end{aligned}
\end{equation}
where $q_\phi\left(\boldsymbol{z} \mid \boldsymbol{X}_u, \boldsymbol{X}_v\right)$ is the latent distribution generated by the encoder, $ p_\theta(\boldsymbol{z} \mid \boldsymbol{X}_v)$ is the prior distribution of neighborhood information, $p_\theta(\boldsymbol{X}_u \mid \boldsymbol{X}_v, \boldsymbol{z})$ is the generative distribution conditioned on $\boldsymbol{z}$ and $\boldsymbol{X}_v$, $\phi$ and $\theta$ represent the variational parameters and generative parameters respectively, $L$ is the number of neighbors of node $v$.
By minimizing the ELBO, the CVAE can learn appropriate parameters for both the encoder and decoder, thus reducing the reconstruction error of the input data under a given condition. 
At the same time, it ensures that the latent distribution generated by the encoder is similar to the prior distribution.

\subsection{Multi-triplet Feature Augmentation for Ponzi Detection} \label{subsec:HLA}
According to Eq.~(\ref{eq: triples}), there exists at least one and at most three different forms of interactions around any target node in HEIG. In order to characterize the triplet information distribution of the target node in a more fine-grained way and generate more diverse triplet-level augmentation features, we perform different triplet-level information generation models specified to types of interactions separately.

Specifically, for each type of target node $v$ (CAs or EOAs), we first obtain its neighbor pairs $(\boldsymbol{X}_v, \boldsymbol{X}_u, u \in \mathcal{N}_v(R^*))$ based on different triplet interaction patterns $R^*$, where
\begin{equation}
  R^* \in 
  \begin{cases}
    \left\{R_\textit{cc}^\textit{c}, R_\textit{cc}^\textit{t}, R_\textit{ce}^\textit{t}\right\} & \text { if } u \in V_\textit{ca} \\ 
    \left\{R_\textit{ec}^\textit{c}, R_\textit{ec}^\textit{t}, R_\textit{ee}^\textit{t}\right\} & \text { if } u \in V_\textit{eoa}
  \end{cases}
\end{equation}
and then use these neighbor pairs to pre-train a CVAE model $f_\text{cvae}$. 
This CVAE model can generate triplet-level augmentation features for specific types of nodes based on specific triplet interaction patterns.
Finally, for CAs or EOAs, we can pre-train three CVAE models and finally generate three types of triplet-level augmentation features:
\begin{equation}
  \begin{aligned}
    \text{For CAs} &:  \hat{\boldsymbol{X}}_\textit{cc}^\textit{c},\  \hat{\boldsymbol{X}}_\textit{cc}^\textit{t},\  \hat{\boldsymbol{X}}_\textit{ce}^\textit{t} \in \mathbb{R} ^ {n_{c}\times d} \\
    \text{For EOAs}&:  \hat{\boldsymbol{X}}_\textit{ec}^\textit{c},\  \hat{\boldsymbol{X}}_\textit{ec}^\textit{t},\  \hat{\boldsymbol{X}}_\textit{ee}^\textit{t} \in \mathbb{R} ^ {n_{e}\times d}
  \end{aligned}
\end{equation}
where $n$ denotes the number of nodes and $d$ denotes the dimension and is of the same length as the initial feature. 

Based on the aforementioned triplet-level augmentation features, we additionally incorporate the initial features $\boldsymbol{X}_c$ and $\boldsymbol{X}_e$ to create integrated feature groups $\mathcal{X}_{c}$ and $\mathcal{X}_{e}$.
Furthermore, in order to obtain comprehensive triplet-level information, we employ a multi-view approach by iteratively repeating the feature augmentation $\mathcal{M}$ times.
Ultimately, the feature groups can be represented as follows:
\begin{equation} \label{eq:7}
    \begin{aligned}
        \mathcal{X}_{c}^\mathcal{M} :&\ \{\hat{\boldsymbol{X}}_\textit{cc}^\textit{c},\  \hat{\boldsymbol{X}}_\textit{cc}^\textit{t},\  \hat{\boldsymbol{X}}_\textit{ce}^\textit{t},\ \boldsymbol{X}_c\}^\mathcal{M}\\
        \mathcal{X}_{e}^\mathcal{M} :&\ \{\hat{\boldsymbol{X}}_\textit{ec}^\textit{c},\  \hat{\boldsymbol{X}}_\textit{ec}^\textit{t},\  \hat{\boldsymbol{X}}_\textit{ee}^\textit{t},\ \boldsymbol{X}_e\}^\mathcal{M}
\end{aligned}
\end{equation}
where $\mathcal{M}$ is the number of views. Moving forward, we commence with the process of training the model to further improve the multi-view triplet-level features.
Essentially, various node types assume distinct roles within the graph. To capture diverse semantic information associated with each node type, we utilize fully connected (FC) layers.
Concurrently, we incorporate an activation function to learn intricate interaction information within the triplet-level augmentation features in Eq.~(\ref{eq:7}). The formulation is depicted as:
\begin{equation} \label{eq:8}
    \begin{aligned}
    \mathbf{h}_i^{\mathcal{M}} = \sigma\ (\boldsymbol{X}_i^{\mathcal{M}}  \cdot \mathbf{\Theta }_{i} )
    \end{aligned}
\end{equation}
where $i\in \{c, e\}$ represents two node types, $\boldsymbol{X}_i^{\mathcal{M}} \in \mathcal{X}_i^\mathcal{M}$ is the matrix in the $i$ feature group, $\mathbf{\Theta }_{i}$ is parameter weight matrix customized for node types. 
And $\mathcal{H}_i^\mathcal{M}$ is the projected feature group specific to the $i$ type, composed of $\mathcal{H}_c^\mathcal{M}:\{\hat{\boldsymbol{h}}_{cc}^c,\hat{\boldsymbol{h}}_{cc}^t,\hat{\boldsymbol{h}}_{ce}^t,\boldsymbol{h}_c\}^\mathcal{M}$ and $\mathcal{H}_e^\mathcal{M}:\{\hat{\boldsymbol{h}}_{ec}^c,\hat{\boldsymbol{h}}_{ec}^t,\hat{\boldsymbol{h}}_{ee}^t,\boldsymbol{h}_c\}^\mathcal{M}$.

After obtaining the projected feature groups, we concatenate the intra-group matrices to obtain the unique projected feature matrix for various types, which are represented as $\hat{ \boldsymbol{h}}_c= [\hat{\boldsymbol{h}}_{cc}^c ||\  \hat{\boldsymbol{h}}_{cc}^t||\  \hat{\boldsymbol{h}}_{ce}^t||\ \boldsymbol{h}_c] \in \mathbb{R}^{n_c\times 4d'}$ 
and $\hat{\boldsymbol{h}}_e= [\hat{\boldsymbol{h}}_\text{ec}^\text{c}||\  \hat{\boldsymbol{h}}_\text{ec}^\text{t}||\  \hat{\boldsymbol{h}}_\text{ee}^\text{t}||\ \boldsymbol{h}_e] \in \mathbb{R}^{n_e\times 4d'}$, where $||$ denotes the concatenation and $d'$ is the dimension of projected features. After joining multi-views information, writing as $\hat{\boldsymbol{h}}_i^\mathcal{M}$.

To obtain a more comprehensive feature representation, we merge various triplet-level features along with initial node features through a concatenation operation. This integration enables the incorporation of multi-triplet information, resulting in a richer and more comprehensive feature representation. 
Nevertheless, problems also arise along with concatenation, which firstly leads to an extension of feature dimensions, making the subsequent learning process require more run-time memory.
Therefore, we use the FC layer to learn the linear relationships in the input features to achieve dimensionality reduction. The formulation is as follows:
\begin{equation}
    \tilde{ \boldsymbol{h}}_i^\mathcal{M} = \hat{\boldsymbol{h}}_i^\mathcal{M} \cdot  \tilde{\mathbf{\Theta }}_{i}
\end{equation}
where  $\tilde{ \boldsymbol{h}}_i^\mathcal{M} \in \mathbb{R}^{n_i\times d''}$ represents the feature matrix obtained after the fully connected layer, incorporates a weighted combination of the input vectors. 
This process effectively integrates each triplet-level augmentation feature.
Then  we use HGNN to process features pertaining to different node types and edge types. For the first layer, we can employ a general HGNN as backbone, e.g. Heterogeneous Graph Transformer (HGT)\cite{hu2020heterogeneous}, denoted as:
\begin{equation}
    \tilde{ \boldsymbol{H}}^\mathcal{M} = \mathcal{F} _{{HGNN}^\mathcal{M}}\ (\tilde{ \boldsymbol{h}}^\mathcal{M}, \mathcal{A} )
\end{equation}
where $\mathcal{F} _{{HGNN}^\mathcal{M}}(\cdot  )$ is the model specified to the $\mathcal{M}^{th}$ view, $\mathcal{A}$ is the heterogeneous adjacent matrix with all triplet types. 

Subsequently, we utilize an average pooling layer to fuse information from multiple views and obtain the holistic representation $ \boldsymbol{H}=MeanPooling(\tilde{\boldsymbol{H}}^1,\tilde{\boldsymbol{H}}^2,\cdots ,\tilde{\boldsymbol{H}}^\mathcal{M})$.
The intermediate representation is then fed into another HGT layer to obtain the final representation:
\begin{equation}
    \boldsymbol{\mathcal{H}} = \mathcal{F} _{HGNN}\ (\boldsymbol{H}, \mathcal{A} )
\end{equation}
where $\boldsymbol{\mathcal{H}}$ is the final representation for the downstream detection task. 

\section{Experiments} \label{sec:exp}

\begin{table*}[t]
\caption{The results of Ponzi scheme detection in terms of Micro-F1(\%) and Standard Deviation(\%). 
Boldface letters are used for the superior result.}
\centering
\resizebox{\linewidth}{!}{
\renewcommand\arraystretch{1.2}
\begin{tabular}{c|c|cc|cccccc|c} 
    \hline\hline
    Datasets                    & \textit{k} & N2V               & MP2V              & GCN               & GAT                & GT                 & RGCN               & HAN               & HGT               & MAHGNN                      \\ 
    \hline
    \multirow{2}{*}{Vesion-0}   & 0.01       & 72.99 $\pm $ 3.01 & 83.64 $\pm $ 4.24 & 84.42 $\pm $ 3.48 & 72.47 $\pm $ 6.54  & 86.23 $\pm $ 2.26  & 88.57 $\pm $ 1.51  & 87.79 $\pm $ 0.64 & 87.79 $\pm $ 1.76 & \textbf{90.91 $\pm $ 2.32}   \\
                                & 0.001      & 74.55 $\pm $ 4.07 & 82.08 $\pm $ 4.68 & 87.27 $\pm $ 3.80 & 74.03 $\pm $ 9.02  & 90.91 $\pm $ 2.72  & 86.23 $\pm $ 1.76  & 87.79 $\pm $ 1.76 & 87.53 $\pm $ 1.76 & \textbf{90.91 $\pm $ 1.61}  \\  
    \hline
    \multirow{2}{*}{Vesion-1}   & 0.01       & 78.96 $\pm $ 2.52 & 83.64 $\pm $ 4.54 & 89.09 $\pm $ 1.76 & 75.32 $\pm $ 5.06  & 89.09 $\pm $ 2.11   & 89.61 $\pm $ 1.64  & 86.75 $\pm $ 2.38 & 89.35 $\pm $ 1.72  & \textbf{91.69 $\pm $ 2.54}   \\
                                & 0.001      & 80.26 $\pm $ 2.23 & 81.30 $\pm $ 3.35 & 86.75 $\pm $ 0.97 & 71.69 $\pm $ 6.22  & 89.61 $\pm $ 2.32  & 90.39 $\pm $ 3.45  & 88.57 $\pm $ 0.52 & 88.57 $\pm $ 0.52 & \textbf{92.99 $\pm $ 1.04}   \\ 
    \hline
    \multirow{2}{*}{Vesion-2}   & 0.01       & 77.14 $\pm $ 3.54 & 83.64 $\pm $ 3.45 & 83.12 $\pm $ 2.17 & 71.95 $\pm $ 3.14  & 83.64 $\pm $ 6.90  & 95.32 $\pm $ 1.32  & 88.83 $\pm $ 1.94 & 90.39 $\pm $ 1.32 & \textbf{95.58 $\pm $ 1.94}   \\
                                & 0.001      & 80.26 $\pm $ 3.01 & 83.90 $\pm $ 1.76 & 83.12 $\pm $ 5.39 & 73.51 $\pm $ 6.01  & 88.05 $\pm $ 3.01  & 92.99 $\pm $ 0.64  & 91.17 $\pm $ 0.52 & 91.43 $\pm $ 0.64 & \textbf{95.58 $\pm $ 1.04}   \\
    \hline\hline
    \end{tabular}}
\label{table:result}
\end{table*}

\subsection{Baselines}

We compare our MAHGNN with various categories of Ponzi detection approaches based on graph representation models, including unsupervised learning (UL)~\cite{barlow1989unsupervised} and semi-supervised learning (SSL)~\cite{zhu2005semi}. The UL category comprises conventional homogeneous graph embedding models and conventional heterogeneous graph embedding models, while the SSL category encompasses GNNs for homogeneous graphs and GNNs for heterogeneous graphs. 
We utilize homogeneous approach for the homogeneous graph associated with the heterogeneous Ethereum interaction network, the baselines are as follows:

\begin{itemize}
    \item \textbf{Node2Vec}~\cite{grover2016node2vec} introduces two parameters, p and q, to Deepwalk~\cite{perozzi2014deepwalk} for regulating the random walk procedure.
    \item \textbf{Metapath2Vec}~\cite{dong2017metapath2vec} creates node sequences by employing meta-paths to limit the sequence of node access, thereby more efficiently capturing the associations between distinct node types in a heterogeneous network. 
    We adopt the well-defined meta-paths in the past work~\cite{jin2022time} and take the optimal result.
    \item \textbf{GCN}~\cite{kipf2016semi} generates a node embedding representation by executing a convolution operation on the graph, merging the node features with those of its adjacent nodes. 
    \item \textbf{GAT}~\cite{velivckovic2017graph} employs an attention mechanism to evaluate the relevance between nodes and their neighbors, thereby more efficiently capturing information regarding the graph structure.
    \item \textbf{GT}~\cite{shi2020masked} advances upon the Transformer~\cite{vaswani2017attention} by converting graph data into a sequence of node vectors, which are subsequently aggregated and interacted with using a multi-layer self-attentive mechanism to acquire a comprehensive representation of the entire graph.
    \item \textbf{RGCN}~\cite{schlichtkrull2018modeling} leverages relationship-specific weight matrices in the convolutional layer to perform node convolution, which is able to handle multiple types of relationships in heterogeneous graphs.
    \item \textbf{HAN}~\cite{wang2019heterogeneous} incorporates the meta-path relationships between different types of entities in the attention mechanism to compute attention coefficients, which enhances the model's ability to capture the dependencies between entities.
    The used meta-path is the same as Metapath2Vec.
    \item \textbf{HGT}~\cite{hu2020heterogeneous} uses the relationships between different types of entities in Graph Transformer to calculate the attention factor to better capture the dependencies between entities.
\end{itemize}
The homogeneous and heterogeneous baselines demonstrate a direct correspondence between them. Specifically, the Node2Vec (N2V) aligns with the Metapath2Vec (MP2V), the GCN aligns with the RGCN, the GAT aligns with the HAN, and the Graph Transformer (GT) aligns with the HGT.

\subsection{Experimental Setup}
In order to better showcase the effectiveness of the augmentation module, we fine-tune the parameters of the baseline methods to their optimal values. 
In the case of the unsupervised random walk method, we set the walk length to 50, window size to 10, walk length per node to 5, and $p$ and $q$ values in Node2Vec to the optimal value are selected from the set \{0.25, 0.4\}.
The hidden dimension and learning rates for the semi-supervised learning methods, including GCN, GAT, GT, RGCN, HAN, HGT and our MAHGNN are selected from sets \{16, 32, 64\} and \{0.01, 0.001\}, respectively. 
Additionally, for RGCN, $num\_bases$ is set to 50.
For all methods involving the multi-head attention, the head is fixed at 4.
For our method, the multi-views parameters $\mathcal{M}$ is chosen from \{1, 2, 3, 4\}, backbone HGNN is HGT, and activation function $\sigma(\cdot)$ is $Tanh(\cdot)$.
For all experiments, we split the dataset into 6:2:2 and report the average Micro-F1 performance after 5 runs.

When working with the Ponzi dataset, training directly on the entire graph can be challenging due to memory limitations. 
To address this, we utilize a neighbor sampling approach to train in smaller batches. 
Our process involves randomly selecting a fixed number of neighbors for each edge type at each layer, for a given target node. 
Then, we allow the node to gather messages from the selected neighbors layer by layer. 
To ensure consistency, we set the number of sampled neighbors to 100 for each edge type at each layer. This sampling strategy is conveniently available in the PyG\footnote[3]{\url{https://pytorch-geometric.readthedocs.io/}} package.
Notably, the pre-training process utilizes the entire graph as it enables the model to undergo separate pre-training based on distinct triplet while keeping memory demands low.

\subsection{Detection Performance}
The results of Ponzi scheme detection are reported in Table~\ref{table:result}, from which we can observe that our method achieves the state-of-the-art performance.  
Our method focuses on obtaining triplet-level augmentation features through extracting multi-triplet information, and then aggregating neighborhood information using a heterogeneous neural network model.
This method consistently achieves best detection performance across various versions of the Ponzi datasets compared to other methods. 
Specifically, the improvement rate of MAHGNN ranges from $0.00\%\thicksim  2.88\%$ compared to the best results of baselines cross different dataset versions. 
Based on the research, it seems that the MAHGNN model is more effective in identifying Ponzi schemes than the HGT model. 
This arises from the MAHGNN model's capacity to generate and gather rich triplet-level structural features.  
This capability facilitates the acquisition of precise structural properties within the graph, ultimately yielding a high-quality representation beneficial for downstream detection tasks.
Additionally, it underscores the significance of discerning these triplet types in heterogeneous graphs when learning graph structure representations.

At the same time, we find that the utilization of heterogeneous methods is indisputably more effective than the utilization of homogeneous methods in detecting Ponzi schemes on most of the various datasets. 
This is because heterogeneous graphs contain an abundance of essential information that enables learning of specific properties of Ponzi interaction behavior. 
Our approach of modeling Ethereum data as Ethereum heterogeneous interaction graphs, while considering the multilateral relationships within these interaction graphs, is thus strongly supported by these findings.

Moreover, upon analyzing the results across different datasets, we discover that the version-2 dataset demonstrates significantly greater effectiveness in detection. 
This observation suggests the existence of crucial structures in the Ethereum interaction network that facilitate the detection of Ponzi schemes. 
Consequently, it is worth considering how to extract these structures from large-scale Ethereum data in future research. 
Additionally, the minimal variation in detection results for different values of \textit{k} highlights the efficacy of our approach to second-order neighbor filtering in preserving the underlying graph structure. 
Furthermore, it also indicates that low transaction volume nodes in the second-order neighborhood provide limited assistance in Ponzi scheme detection.

\section{Conclusion} \label{sec:Conclusion}
The current detection model for Ethereum Ponzi schemes primarily relies on homogeneous design. 
While existing heterogeneous models show power in detecting Ponzi schemes, they still lack efficient heterogeneous graph augmentation strategies. 
To address this limitation, we propose the MAHGNN for Ponzi scheme detection. In detail, our approach leverages the presence of multi-triplet interactions within the HIG. Then we pre-train a CVAE model for different triplet types to capture individual triplet-level structure information. 
Experimental results demonstrate the effectiveness of our approach, which is better than existing detection model. However, considering the time-consuming nature of the pre-training process, we are motivated to explore an end-to-end framework that can achieve more efficient detection of Ponzi schemes in the future.

\bibliographystyle{IEEEtran} 
\bibliography{mybibliography,IEEEabrv}

\end{document}